\documentclass[argument]{aastex61}
\submitjournal{ApJL}

\begin{document}

\title{A solar blowout jet caused by the eruption of a magnetic flux rope}

\correspondingauthor{Xiaoshuai Zhu}
\email{xszhu@bao.a.cn}

\author{Xiaoshuai Zhu}
\affiliation{Key Laboratory of Solar Activity, National Astronomical Observatories, Chinese Academy of Sciences; \color{blue}{xszhu@bao.ac.cn}}
\affiliation{Key Laboratory of Modern Astronomy and Astrophysics (Nanjing University), Ministry of Education, Nanjing 210093, China}

\author{Huaning Wang}
\affiliation{Key Laboratory of Solar Activity, National Astronomical Observatories, Chinese Academy of Sciences; \color{blue}{xszhu@bao.ac.cn}}
\affiliation{University of Chinese Academy of Sciences, China; \color{blue}{hnwang@nao.cas.cn}}

\author{Xin Cheng}
\affiliation{Nanjing University, China; \color{blue}{xincheng@nju.edu.cn}}
\affiliation{Key Laboratory of Modern Astronomy and Astrophysics (Nanjing University), Ministry of Education, Nanjing 210093, China}

\author{Chong Huang}
\affiliation{Institute of Space Sciences and School of Space Science and Physics, Shandong University, Weihai 264209, China}
\affiliation{Key Laboratory of Solar Activity, National Astronomical Observatories, Chinese Academy of Sciences; \color{blue}{xszhu@bao.ac.cn}}



\begin{abstract}
We investigate the three-dimensional (3D) magnetic structure of a blowout jet originated in the west edge of NOAA Active Region (AR) 11513 on 02 July 2012 by means of recently developed forced field extrapolation (FFE) model. The results show that the blowout jet was caused by the eruption of the magnetic flux rope (MFR) consisting of twisted field lines. We further calculate the twist number $\mathcal{T}_{w}$ and squashing factor Q of the reconstructed magnetic field and find that (1) the MFR corresponds well to the high $\mathcal{T}_{w}$ region (2) the MFR outer boundary corresponds well to the high Q region, probably interpreting the bright structure at the base of the jet. The twist number of the MFR is estimated to be $\mathcal{T}_{w}=-1.54\pm 0.67$. Thus, the kink instability is regarded as the initiation mechanism of the blowout jet as $\mathcal{T}_{w}$ reaching or even exceeding the threshold value of the kink instability. Our results also indicate that the bright point at the decaying phase is actually comprised of some small loops that are heated by the reconnection occurred above. In summary, the blowout jet is mostly consistent with the scenario proposed by \citet{mcs10} except that the kink instability is found to be a possible trigger.

\end{abstract}




\section{Introduction} \label{sec:intro}
The concept of ``blowout jet'' was first introduced by \citet{mcs10} based on the morphological description of X-ray jet in the \textit{Hinode}/X-ray Telescope movie. The broad spire and bright base arch of the ``blowout jet'' distinguish from the thin spire and relatively dim base arch of standard jet. In the widely accepted scenario of a blowout jet \citep{mcs10,rpp16}: the sheared or twisted arch field is supposed to emerge from below the photosphere, forming a current sheet at the interface between the arch field and ambient open field. The onset of magnetic reconnection at the current sheet shows the similar feature with the standard jet apparently, then the sheared or twisted arch field is erupted outward as the key structure of a CME \citep{c11}. Among the total number of 109 jets examined in \citet{mcs10} and \citet{msf13}, 50 are blowout, 53 are standard, 6 are ambiguous.

A minifilament whose magnetic structure is argued to be a helical MFR, is often observed at the base of the blowout jet \citep{hjz11,hjy13,sls12,asm14, smf15}. The existence of MFR is supported by the helical structure in the spire during the untwisting motion of the erupting mass \citep[e.g.,][]{ppv08,nbp09,sls11,ctk12,czm12,mse12,sls12,lim13,zj14,lwl14,cpt15}. In addition, 3D magnetohydrodynamic (MHD) jet models \citep{pad09,pad10,pdd15,rpd10,cpt15,kda17} according to the eruption of the twisted magnetic field also show the same helical motion as the observations.

Magnetic structures of the source regions of the blowout jets have been modeled using potential \citep{ldl11,zcg12}, linear \citep{mgu08} and non-linear force-free modelings \citep{sgm13,gds13}. The previous works just unveiled the weakly sheared core fields and opened ambient fields. The twisted MFRs in which more magnetic free energy is stored to power the blowout jets, however, have never been disclosed.

In this letter, we report an MFR at the base of a blowout jet. The MFR is successfully reconstructed by recently developed FFE model \citep{zwd13,zwd16}. To study the change of the magnetic field, we make a time series of extrapolations using Helioseismic and Magnetic Imager \citep[HMI;][]{ssb12,hlh14} vector magnetograms. We further calculate the twist number \citep{bp06} and squashing factor \citep{dpm96,thd02,pd12} to study the property of the MFR. The observational data sets are described in Section \ref{sec:data}, the evolution of the blowout jet is presented in Section \ref{sec:evolution}, the extrapolation results are analyzed in Section \ref{sec:extrapolation}, which is followed by the discussions and conclusions in Section \ref{sec:discussion}.

\section{Observational data} \label{sec:data}

\begin{figure*}
\plotone{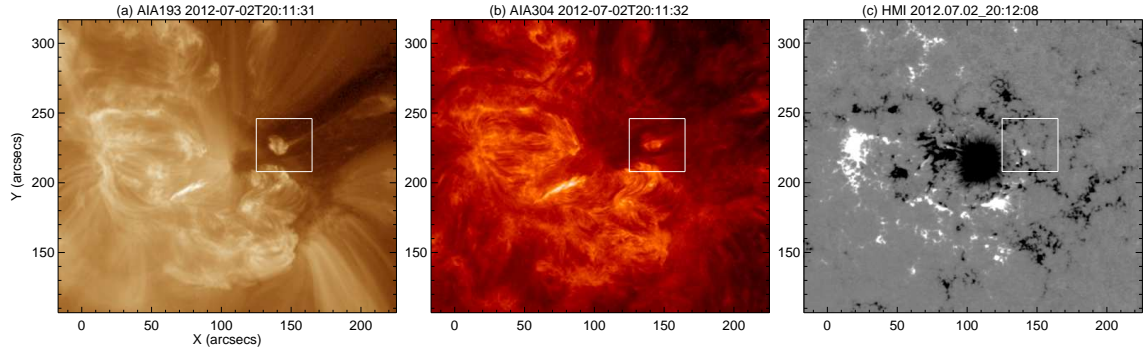}
\caption{SDO/AIA 193$\AA$ (a), 304 $\AA$ (b) images and line of sight (LOS) magnetogram of AR 11513. The white box shows the location of the jet.
\label{fig:AR11513}}
\end{figure*}

Recurrent jets were observed at the boundary between the AR 11513 and the neighboring coronal hole on July 2, 2012 (see white box in Figure \ref{fig:AR11513}). Our attention is paid to the blowout jet occurred at 21:12UT.

HMI onboard the Solar Dynamics Observatory \citep[SDO;][]{ptc12} provides 45 seconds line of sight (LOS) magnetograms and 12 minutes vector magnetograms. Both of their pixel size is $0.5''$. The Atmospheric Imaging Assembly \citep[AIA;][]{lta12} also on board SDO provides full disk images of solar corona at multiple EUV passbands with cadence of 12 seconds and pixel size of $0.6''$. We also used the H$\alpha$ data observed at Big Bear Solar Observatories (BBSO) to study the evolution of the jet in the chromosphere.

\section{Evolution of the blowout jet} \label{sec:evolution}

\begin{figure*}
\plotone{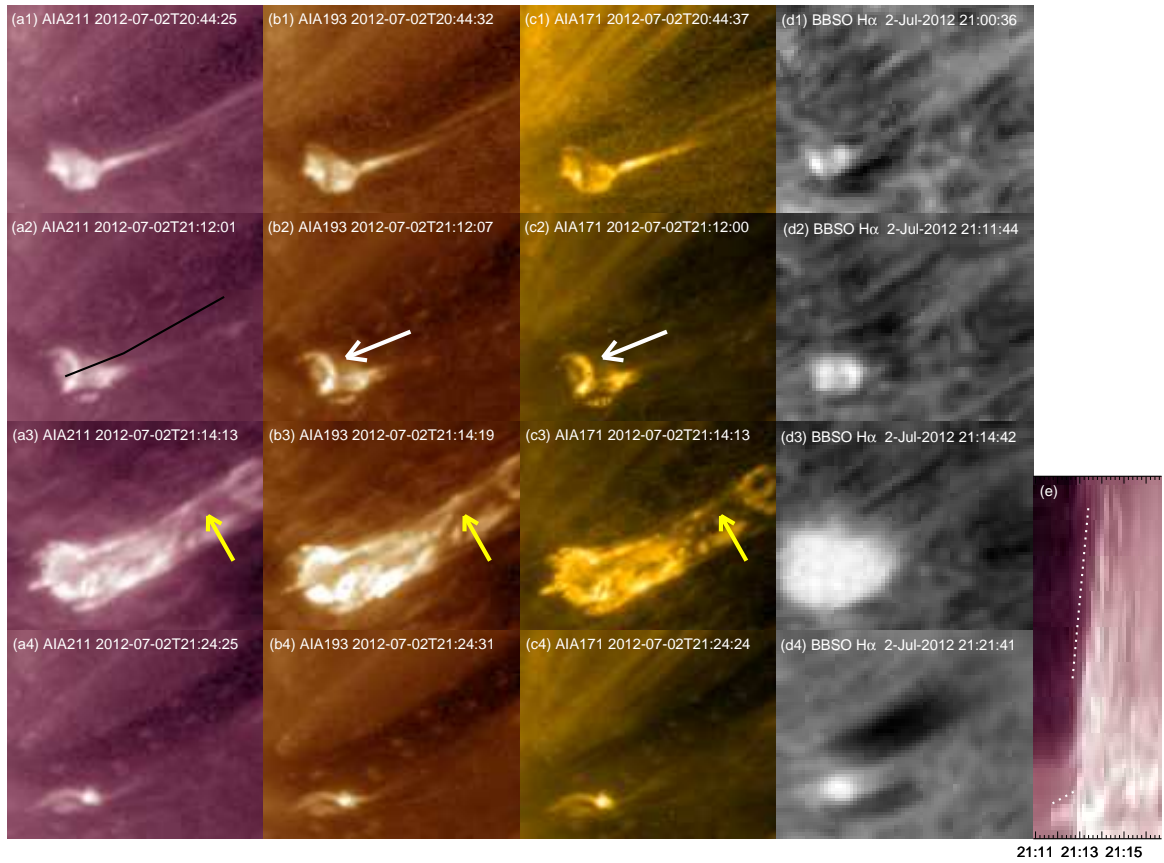}
\caption{SDO/AIA 211 $\AA$, 193 $\AA$, 171 $\AA$, BBSO H$\alpha$ show the evolution of the blowout jet. The arrows show: (b2, c2) the bright tube; and (a3, b3, c3) the helical structure. (e): Distance-time plot of black line in (a2). (Animation of this figure is available.)
\label{fig:evolution}}
\end{figure*}

Figure \ref{fig:evolution} and online movie show the blowout jet in different passbands. Here, the evolution is divided into three stages.

First stage: before 21:11, the jet's base appeared as a circular shape (Figure \ref{fig:evolution} (a1, b1, c1, d1)) which could be the combination of several dipoles, the loops connect which may be heated. The plasma is observed to intermittently move out along the open field lines even though the whole structure is stable.

Second stage: at 21:11, a bright point at the south of the circular area appeared and then quickly extended to the north to form a bent tube (pointed by the white arrows in Figure \ref{fig:evolution} (b2, c2)). The tube increasingly got brightened, followed by a slowly upward motion (lower dot line in Figure \ref{fig:evolution} (e)) and a fast ejection motion (upper dot line in Figure \ref{fig:evolution} (e)). The strong brightening of the jet's base suggests that the internal reconnection occurs between the opposite-polarity stretched legs of the erupting structure. Meanwhile, the jet's spire shows multi-stranded curtain structure with rotating motion in the broaden spire (pointed by the yellow arrows in Figure \ref{fig:evolution} (a3, b3, c3)). All these are typical morphological characteristics of a blowout jet.

Third stage: at 21:18, the jet's base and spire started to decay. All bright structure faded away except a dimming bright point (Figure \ref{fig:evolution} (a4, b4, c4, d4)).

\section{3D magnetic structure of the blowout jet} \label{sec:extrapolation}

To understand the magnetic structure and evolution of the jet, we use the FFE model that utilizes the MHD relaxation method (full MHD equations are solved) to build the equilibrium state of the system that approximate the solar atmosphere. The HMI vector magnetograms are taken as bottom boundary condition. The FFE model is particularly suited to compute the magnetic field in the chromosphere, transition region and low corona because of the relatively high plasma $\beta$ there. It has been successfully used to reproduce the magnetic structure of H$\alpha$ fibrils \citep{zwd16}, small filament \citep{wlz16}, and bright arcade in the chromosphere or low corona \citep{zsl17}. In the work, the extrapolation is performed in the cubic box resolved by 480*416*128 grid points with $\triangle x=\triangle y=\triangle z=0.5''$. The photosphere boundary field of view for extrapolation is shown in Figure \ref{fig:AR11513}.

\subsection{The evolution of magnetic structure}

\begin{figure*}
\plotone{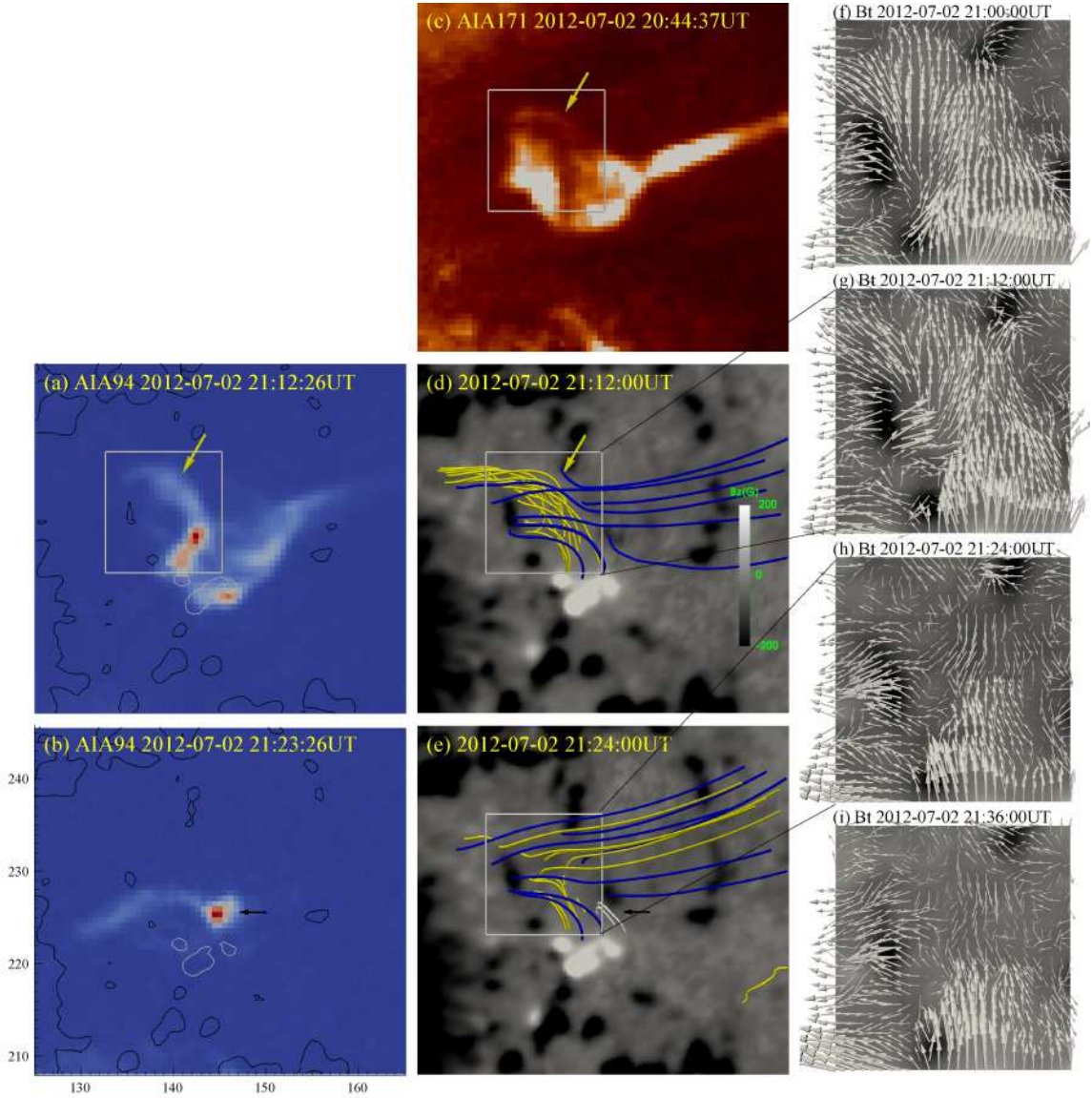}
\caption{SDO/AIA 171 94 $\AA$ images, magnetic field lines and vector magnetograms show the state before and after the jet. The yellow arrows in the panel (a) and (d) point to the bright tube and MFR, respectively. The contour in the panel (a, b) in black/white color represents the LOS magnetic field strength of -200G/200G. The yellow/blue lines in the panel (d) outline the MFR/ambient magnetic field. The white field lines in the (e) show the small loops at the post jet bright point pointed by the black arrow in the panel (b) and (e).
\label{fig:structure}}
\end{figure*}

Figure \ref{fig:structure} (d) shows that an MFR (yellow lines) appears at the source region of the jet. The MFR corresponds well to the observed bright tube (Figure \ref{fig:structure} (a)) as seen in AIA images. With the jet eruption, most of the twisted lines are released, only leaving some untwisted and open field lines in place (Figure \ref{fig:structure} (e)). The small loops (white field lines in panel (e)) at the bright point (Figure \ref{fig:structure} (b)) are possibly the reconnected field lines. Although we can not see the dynamic process of the jet by extrapolation, the change of the magnetic field clearly display that the MFR is ejected during the jet.

The arrows in Figure \ref{fig:structure} (g) show the transverse field which is aligned with the MFR. After the jet, the transverse field decreases and becomes disordered (Figure \ref{fig:structure} (h,i)). This is consistent with the fact that the eruption of the jet takes away most of twisted field and just leaves some small closed field lines and large-scale open field.

\subsection{The structure of MFR}

\citet{bp06} defined the twist of the neighboring magnetic field lines, which is related to the parallel electric current ($J_{\|}$), as follows:
\begin{eqnarray}
\mathcal{T}_{w}&=&\int_{s} \frac{\mu_{0}J_{\|}}{4\pi|\mathbf{B}|}ds \nonumber \\
&=&\int_{s} \frac{(\nabla \times \mathbf{B})\cdot \mathbf{B}}{4\pi B^{2}}ds, \nonumber
\end{eqnarray}
where the integration is carried out along the specific field line.

\citet{dpm96} introduced the quasi-separatrix layers (QSLs) as the generalized topological structure. The QSLs are defined by high squashing factor Q regions where the connection of the magnetic field varies strongly. Q is defined by mapping the field line \citep{thd02}:
\begin{eqnarray}
D_{12} & = & \left(
\begin{array}{cc}
\partial x_2/\partial x_1 \quad \partial x_2/\partial y_1\\
\partial y_2/\partial x_1 \quad \partial y_2/\partial y_1 \nonumber
\end{array}\right)\\
&=&  \left(
\begin{array}{cc}
a \qquad b \\
c \qquad d
\end{array}\right),\nonumber \\
Q  &=&  \frac{a^2+b^2+c^2+d^2}{|B_n(x_1,y_1)/B_n(x_2,y_2)|},\nonumber
\end{eqnarray}
where ($x_1$, $y_1$) and ($x_2$, $y_2$) are the two footpoints of a field line.

The code we used to calculate the twist number $\mathcal{T}_{w}$ and squashing factor Q is developed by \citet{lkt16}. To save computation resource, we select sub-domain $x\in[130.0, 160.3], y\in[231.0, 240.3]$ and $z\in[0.0, 10.1]$ where x (+x towards west) and y (+y towards north) are the heliocentric coordinate and z is the height. The sub-domain was resolved by 960*880*320 grids when computing $\mathcal{T}_{w}$ and Q. Therefore, the grids are refined by 16 times after extrapolation.

\begin{figure*}
\plotone{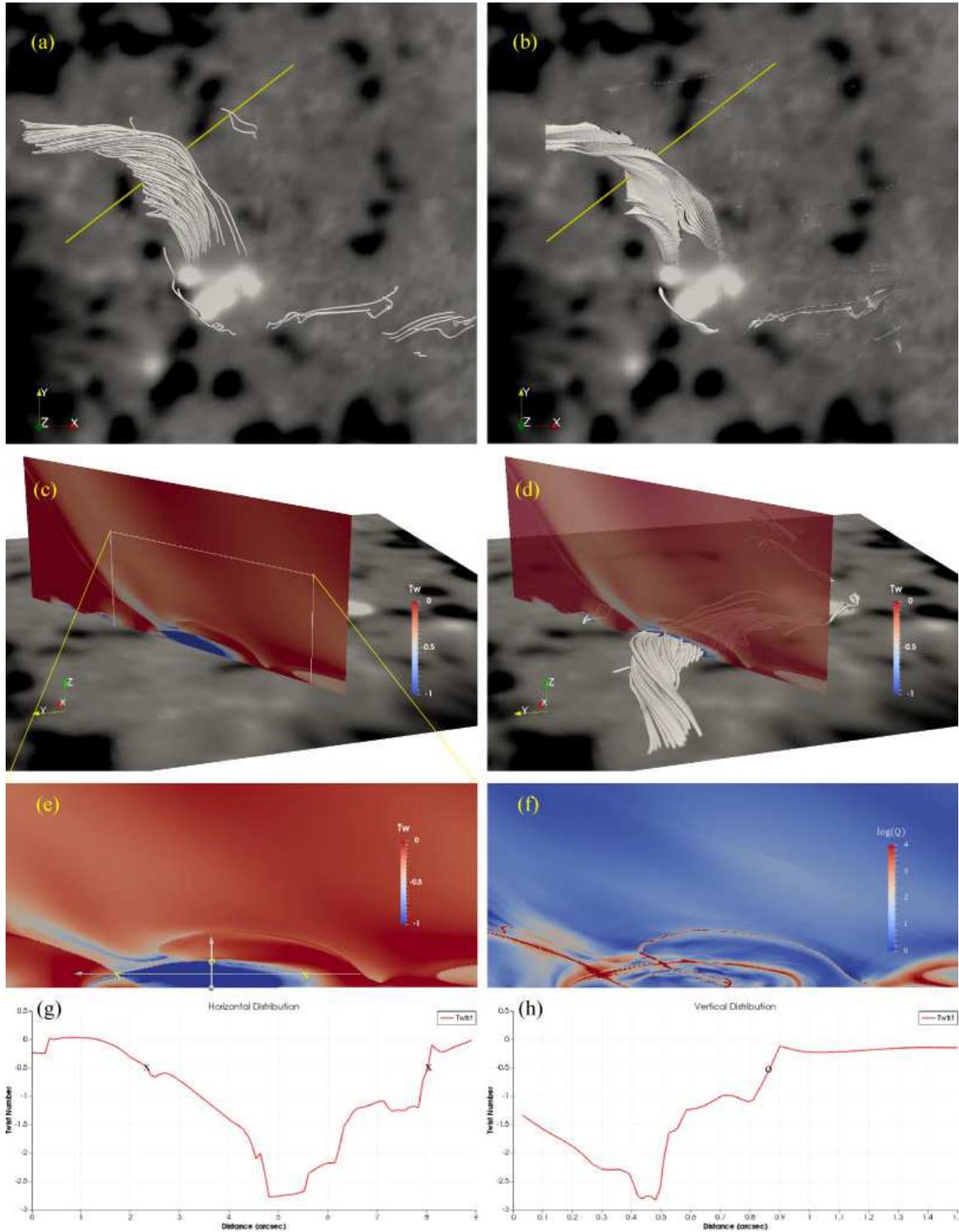}
\caption{Extrapolated 3D magnetic structure (a) and 3D contour of $\mathcal{T}_{w}=-1.5$ (b) of the MFR. (c) The twist number in the cutting plane (denoted by yellow line in the panel (a)). (d) The MFR inside the boundary of $\mathcal{T}_{w}=-0.5$. (e,f) $\mathcal{T}_{w}$ and Q in the magnified cutting plane. (g,h) $\mathcal{T}_{w}$ distribution along horizontal and vertical direction. ``x'' and ``o'' in panels (e,g,h) indicate the boundary of the MFR. The length and height of the MFR is about 5.5 arcsec 0.9 arcsec, respectively.
\label{fig:mfr}}
\end{figure*}

Figure \ref{fig:mfr}(a) shows the extrapolated 3D field lines of the MFR. The contour of $\mathcal{T}_{w}=-1.5$ (see (b)) marks the MFR accurately. Figure \ref{fig:mfr} (e) and (f) shows a 2D plane of $\mathcal{T}_{w}$ and Q perpendicular to the axis of the MFR. We can see that the $\mathcal{T}_{w}$ has a sharp edge which is consistent with the regions of high Q value. In an MFR, field lines winding around an axis have similar connectivity. QSLs separate the twisted field lines from ambient field lines, which are typical features of an active-region-scaled MFR \citep[e.g.,][]{thd02,gdc13,cdz14,lkt16}. Assuming $\mathcal{T}_{w}=-0.5$ as the boundary, field lines inside have a twist number of $\mathcal{T}_{w}=-1.54\pm 0.67$. The twist at the center of the MFR exceeds to 2.0 turns while decreases to 0.5 toward the edge. \citet{tk03} shows that the \citet{td99} MFR is kink unstable for $|\mathcal{T}_{w}|>1.75$ with aspect ratio $R/r=5$ (R and r are the major and minor radius of the MFR). The instability threshold decrease with decreasing aspect ratio \citep{tk03}. Assuming the length (21 arcsec) and width (5.5 arcsec, see Figure \ref{fig:mfr} (g)) of the extrapolated MFR approximate the major and minor diameters, respectively. The aspect ratio is estimated to be 3.8, implying a smaller kink-instability threshold than 1.75 turns. Therefore, the small scale MFR may be marginally kink unstable. The decay index of the magnetic field above the MFR is about 0.3, which means that the MFR is far below the height where torus instability will occur (the critical decay index requires to be 1.5, \citet{kt06}).

\subsection{Noise and change of the magnetic field on the photosphere}

\begin{figure*}
\plotone{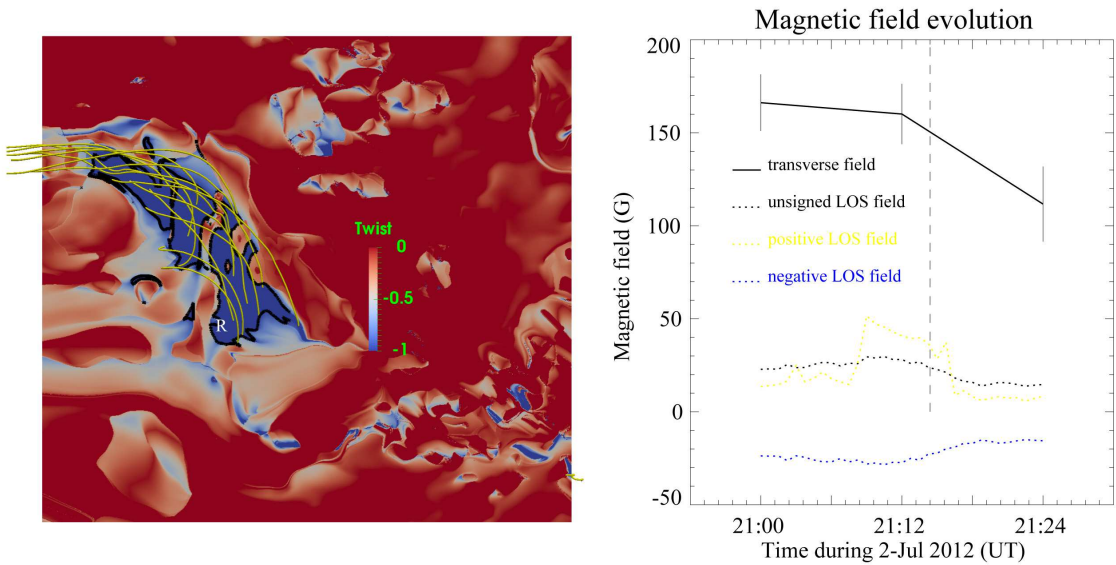}
\caption{Temporal profile of the mean magnetic field on the photosphere. Left: MFR field lines (the same with the yellow lines in Figure \ref{fig:structure} (d)) with the background shows the $\mathcal{T}_{w}$ distribution. The black curve outlines the region (labeled by ``R'') where the magnetic field has strong twist $\mathcal{T}_{w}<-1.0$. Right: the temporal profile of the mean positive, negative and unsigned LOS field and transverse field of the region ``R'' in 24 minutes. The cadence of the two data sets are 45 seconds and 12 minutes, respectively.
\label{fig:fig5-change}}
\end{figure*}

The noise of the transverse magnetic field is large in weak field region because of the nonlinear dependence between the linear polarization and field strength. This lead to unreliable vector magnetic field inversion in solar quiet regions. The jet we analyzed occurred at the boundary of an AR and a coronal hole, which is the interface area of the strong and weak magnetic field. Therefore, it is necessary to assess the noise of the transverse magnetic field at the jet source region. The SDO/HMI provides the standard deviation of inverted magnetic field with data segments $\_$ERR. For example, FIELD$\_$ERR and INCLINATION$\_$ERR are the standard deviation of field strength and inclination angle relative to the LOS. Hence, it is convenient to compute the uncertainty of the transverse field. The temporal profile of the magnetic field is showed in Figure \ref{fig:fig5-change}. Typically at 21:12 UT, the average LOS field, average transverse field, average noise of the transverse field, and the average signal to noise of region ``R'' (surrounded by black curves in Figure \ref{fig:fig5-change} left): are: 28G, 160G, 32G, and 5.4, respectively. The transverse field on the photosphere is about 5.7 times larger than the LOS field under the MFR, which indicate the field lines at this area are nearly horizontal. This results in the relatively small noise of transverse field. The uncertainty of the transverse field is about 18\%, 20\%, and 36\% at 21:00:00, 21:12:00, and 21:24:00, respectively (see the error bar in Figure \ref{fig:fig5-change} right). The high signal to noise of the data denotes it could be used in extrapolation.

The largely different field configuration mainly results from the change of transverse magnetic field on the photosphere after the jet took place. The region ``R'' has a pronounced, 30\% decrease of the transverse field (solid line in Figure \ref{fig:fig5-change} right) from 160G at 21:12:00 before the jet to 112G at 21:24:00 after the jet in 12 minutes. Figure \ref{fig:structure} (f-i) also show decrease and less sheared of the transverse field after the jet. The decrease of the positive, negative, and unsigned LOS field (right panel of Figure \ref{fig:fig5-change}) suggest that the flux cancelation took place at the jet's source region.

\section{Discussion and Conclusion} \label{sec:discussion}

A blowout jet was observed on 2 July 2012 at the west edge of AR 11513. In a previous paper, \citet{csy15} suggested that the rotation and shear motion of the magnetic field build up the free energy to make the jet blow out. In the current work, we further study the 3D magnetic structure of the jet's source region by recently developed FFE model. The twist number and squashing factor are calculated to analyze the magnetic property of this jet. We have the following findings:

First, the transverse magnetic field decreased during the jet. The originally twisted and closed field lines are released, manifesting as the bright base and broaden helical spire, finally just leaving some untwisted and opened field lines in place.

Second, an MFR, reconstructed by the FEE method and being cospatial with the bright tube, is found to exist before the jet and then disappear after the jet blows out. A sharp boundary of the MFR can be seen at 2D cutting plane of $\mathcal{T}_{w}$ distribution. This boundary also corresponds well with the layer with very high Q value that distinguishes the twisted field lines of the MFR from outside.

Third, the twist number of the MFR is $\mathcal{T}_{w}=-1.54\pm 0.67$ with the small aspect ratio $R/r=3.8$, which indicates that the blowout jet is likely triggered by kink instability. The low decay index prevents the eruption from torus instability. 

Combining the observed features with reconstructed 3D magnetic structures, we can argue that before the onset of the blowout jet, a highly twisted MFR exists at the source region of the jet. The twist of the MFR may continuously increase because of the plasma motion or magnetic cancelation at the photosphere. Homologous jets erupted before the blowout one remove the restraining overly field lines. When the twist exceeds a critical value, kink instability takes place and leads to the MFR being ejected. As the MFR moving upward, the internal reconnection occurs between the stretched field lines below. The reconnection outflows may take on a bright core of the jet. Meanwhile, the eruption of the heated MFR in the partly opened ambient field shows multi-strand curtain structure. The helical motion observed in the spire indicates the untwisting process of the erupting MFR. Finally, the jet's base gradually fade away with a weak bright point. This bright point may denote small loops that are heated by the reconnection above. In short, the process of blowout jet is mostly consistent with the scenario proposed by \citet{mcs10}, except that the kink instability is considered to be its initiation mechanism. It has to be pointed out that the direct observation of the twist, for example the twist between fine structures of a filament \citep{wcl15}, is a stronger and direct piece of evidence for the MFR existence. In the future, more case studies, even a statistical study, of 3D magnetic structures of blowout jets will be presented.

\acknowledgments

The authors thank the referee for constructive suggestions. This work is jointly supported by National Natural Science Foundation of China (NSFC) through grants 11403044, 11673035 and 11273031; collaborating Research Program of CAS Key Laboratory of Solar Activity, National Astronomical Observatories(KLSA201609); Natural Science Foundation of Shandong Province (ZR2014AP010). The data used are courtesy of NASA/SDO and the AIA and HMI science teams. The BBSO operation is supported by NJIT, US NSF AGS-1250818, and NASA NNX13AG14G grants.

%




\begin{thebibliography}{}
\bibitem[Antiochos et al.(1999)]{adk99} Antiochos, S. K., DeVore, C. R., Klimchuk, J. A. \ 1999, \apj, 510, 485
\bibitem[Adams et al.(2014)]{asm14} Adams, M., Sterling, Alphonse C., Moore, Ronald L., et al. \ 2014, \apj, 783, 11
\bibitem[Berger \& Prior(2006)]{bp06} Berger, M. A., \& Prior, C.\ 2006, Journal of Physics A: Mathematical and Theoretical, 39, 26
\bibitem[Chen (2011)]{c11} Chen, P. F.\ 2011, Living Reviews in Solar Physics, 8, 1
\bibitem[Chen et al.(2015)]{csy15} Chen, J., Su, J., Yin, Z., et al.\ 2015, \apj, 815, 71
\bibitem[Chen et al.(2012)]{czm12} Chen, H.D., Zhang, J., Ma, S.L.\ 2012, Research in Astronomy and Astrophysics, 12, 573
\bibitem[Cheng et al.(2014)]{cdz14} Cheng, X., Ding, M.D., Zhang, J., et al.\ 2014, \apj, 789, 93
\bibitem[Cheung et al.(2015)]{cpt15} Cheung, Mark C. M., De Pontieu, B., Tarbell, T. D., et al.\ 2015, \apj, 801, 83
\bibitem[Curdt et al.(2012)]{ctk12} Curdt, W.£¬ Tian, H.£¬ Kamio, S.\ 2012, \solphys, 280, 417
\bibitem[D\'{e}moulin et al.(1996)]{dpm96} D\'{e}moulin, P., Prest, E. R., Mandrini, C. H.\ 1996, \aap, 308, 643
\bibitem[Guo et al.(2013)]{gdc13} Guo, Y., Ding, M. D., Cheng, X., et al.\ 2013, \apj, 779, 157
\bibitem[Guo et al.(2013)]{gds13} Guo, Y., D\'{e}moulin, P., Schmieder, B., et al.\ 2013, \aap, 555, 19
\bibitem[Hong et al.(2013)]{hjy13} Hong, J., Jiang, Y., Yang, J., et al.\ 2013, Research in Astronomy and Astrophysics, 13, 253
\bibitem[Hong et al.(2011)]{hjz11} Hong, J., Jiang, Y., Zheng, R., et al.\ 2011, \apjl, 738, 20
\bibitem[Hoeksema et al.(2014)]{hlh14} Hoeksema, J. T., Liu, Y., Hayashi, K., et al.\ 2014, \solphys, 289, 3483
\bibitem[Karpen et al.(2017)]{kda17}  Karpen, J. T., DeVore, C. R., Antiochos, S. K., et al.\ 2017, \apj, 834, 62
\bibitem[Kliem \& T\"{o}r\"{o}k(2006)]{kt06}  Kliem, B., \& T\"{o}r\"{o}k, T.\ 2006, \prl, 96, 25502
\bibitem[Lee et al.(2013)]{lim13} Lee, K.-S., Innes, D. E., Moon, Y.-J., et al.\ 2013, \apjl, 766, 1
\bibitem[Lemen et al.(2012)]{lta12} Lemen, J. R., Title, A. M., Akin, D. J., et al.\ 2012, \solphys, 275, 17
\bibitem[Liu et al.(2011)]{ldl11} Liu, C., Deng, N., Liu, R., et al.\ 2011, \apjl, 735, 18
\bibitem[Liu et al.(2014)]{lwl14} Liu, J., Wang, Y., Liu, R., et al.\ 2014, \apjl, 782, 94
\bibitem[Liu et al.(2016)]{lkt16} Liu, R., Kliem, B., Titov, Viacheslav S., et al.\ 2016, \apj, 818, 148
\bibitem[Moore et al.(2010)]{mcs10} Moore, Ronald L., Cirtain, Jonathan W., Sterling, Alphonse C., et al.\ 2010, \apj, 720, 757
\bibitem[Moore et al.(2013)]{msf13} Moore, Ronald L., Sterling, Alphonse C., Falconer, David A., et al.\ 2013, \apj, 769, 134
\bibitem[Moreno-Insertis et al.(2008)]{mgu08} Moreno-Insertis, F., Galsgaard, K., Ugarte-Urra, I.\ 2008, \apjl, 673, 211
\bibitem[Morton et al.(2012)]{mse12} Morton, R. J., Srivastava, A. K., Erd¨¦lyi, R., et al.\ 2012, \aap, 542, 70
\bibitem[Nistic¨° et al.(2009)]{nbp09} Nistic¨°, G., Bothmer, V., Patsourakos, S., et al.\ 2009, \solphys, 259, 87
\bibitem[Pariat et al.(2009)]{pad09} Pariat, E., Antiochos, S. K., DeVore, C. R.\ 2009, \apj, 691, 61
\bibitem[Pariat et al.(2010)]{pad10} Pariat, E., Antiochos, S. K., DeVore, C. R.\ 2010, \apj, 714, 1762
\bibitem[Pariat et al.(2012)]{pd12} Pariat, E., \& D\'{e}moulin, P.\ 2012, \aap, 541, A78
\bibitem[Pariat et al.(2015)]{pdd15} Pariat, E., Dalmasse, K., DeVore C. R.\ 2015, \aap, 573, 130
\bibitem[Patsourakos et al.(2008)]{ppv08} Patsourakos, S., Pariat, E., Vourlidas, A., et al.\ 2008, \apjl, 680, 73
\bibitem[Pesnell et al.(2012)]{ptc12} Pesnell, W. D., Thompson, B. J., Chamberlin, P. C.\ 2012, \solphys, 275, 3
\bibitem[Rachmeler et al.(2010)]{rpd10} Rachmeler, L. A., Pariat, E., DeForest, C. E., et al.\ 2010, \apj, 715, 1556
\bibitem[Raouafi et al.(2016)]{rpp16} Raouafi, N. E., Patsourakos, S., Pariat, E., et al.\ 2016, \ssr, 201, 1
\bibitem[Schmieder et al.(2013)]{sgm13} Schmieder, B., Guo, Y., Moreno-Insertis, et al.\ 2013, \aap, 559, 1
\bibitem[Schou et al.(2012)]{ssb12} Schou, J., Scherrer, P. H., Bush, R. I., et al.\ 2012, \solphys, 275, 229
\bibitem[Shen et al.(2011)]{sls11} Shen, Y., Liu, Y., Su, J., et al.\ 2011, \apjl, 735, 43
\bibitem[Shen et al.(2012)]{sls12} Shen, Y., Liu, Y., Su, J., et al.\ 2012, \apj, 745, 164
\bibitem[Sterling et al.(2015)]{smf15} Sterling, Alphonse C., Moore, Ronald L., Falconer, David A., et al.\ 2015, \nat, 523, 437
\bibitem[Titov \& D\'{e}moulin (1999)]{td99} Titov, V. S., \& D\'{e}moulin, P.\ 1999, \aap, 406, 1043
\bibitem[Titov et al.(2002)]{thd02} Titov, V. S., Hornig, G., \& D\'{e}moulin, P.\ 2002, \jgr, 107, 1164
\bibitem[T\"{o}r\"{o}k \& Kliem(2003)]{tk03} T\"{o}r\"{o}k, T., \& Kliem, B.\ 2003, \aap, 406, 1043
\bibitem[Wang et al.(2015)]{wcl15} Wang, H., Cao, W., Liu, C., et al.\ 2015, Nature Communications, 6, 7008
\bibitem[Wang et al.(2016)]{wlz16} Wang, R., Liu, Y., Zimovet, I., et al.\ 2016, \apjl, 827, 12
\bibitem[Zhang et al.(2012)]{zcg12} Zhang, Q. M., Chen, P. F., Guo, Y.\ 2012, \apj, 746, 19
\bibitem[Zhang \& Ji(2014)]{zj14} Zhang, Q. M., \& Ji, H. S.\ 2014, \aap, 561, 134
\bibitem[Zhao et al.(2017)]{zsl17} Zhao, J., Schmieder, B., Li, H., et al.\ 2017, \apj, 836, 52
\bibitem[Zhu et al.(2013)]{zwd13} Zhu, X., Wang H., Du Z., et al.\ 2013, \apj, 768, 119
\bibitem[Zhu et al.(2016)]{zwd16} Zhu, X., Wang H., Du Z., et al.\ 2016, \apj, 826, 51
\end{thebibliography}
\end{document}